\documentclass[twocolumn, floatfix,showpacs,preprintnumbers,amsmath,amssymb]{jpsj3}
\usepackage{graphicx}
\usepackage{pifont}
\graphicspath{{fig/}}
\usepackage{dcolumn}
\usepackage{bm}
\usepackage{float}
\usepackage{amsmath,amssymb}
\usepackage{color}
\usepackage[T1]{fontenc}
\usepackage{textcomp}

\title{ Magnon Pumping by a Time-Dependent Transverse Magnetic Field in Ferromagnetic Insulators }

\author{
Kouki Nakata\thanks{E-mail address: nakata@yukawa.kyoto-u.ac.jp} and Gen Tatara 
}

\inst{ $^*$Yukawa Institute for Theoretical Physics, 
Kyoto University, Kitashirakawa Oiwake-Cho, Kyoto 606-8502, Japan

Department of Physics, Tokyo Metropolitan University, Hachioji, Tokyo 192-0397, Japan
}

\abst{
The magnon pumping effect in ferromagnetic insulators under an external time-dependent transverse magnetic field
is theoretically studied.
Generation of a magnon current is discussed by calculating the magnon source term in the spin continuity equation.
This term represents the non-conservation of magnons arising from an applied transverse magnetic field.
The magnon source term has a resonance structure as a function of the angular frequency of the transverse field,
and this fact is useful to enhance the pumping effect.
}

\kword{magnon pumping, magnon current, magnon source term, spintronics, resonance}

\begin{document}
\maketitle

\section{Introduction}
\label{sec:intro}

Recently a new branch of physics and nanotechnology called spintronics\cite{maekawa,mod,mod2} has emerged 
and has been attaching special attention from viewpoints of the fundamental science and application. 
The aim of spintronics is the control of 
the spin as well as charge degrees of freedom of electrons, 
and therefore  establishing methods for generation and observation of a spin current is an urgent issue.  

A standard way to generate a spin current is the spin pumping\cite{takeuchi,mizukami, pump2,pump3,referee} 
effect in ferromagnetic-normal metal junctions. 
There the  precession  of  the  magnetization caused by an external field 
induces a spin current pumped into a normal metal.  
This method was  theoretically proposed by R.H.Silsbee et.al\cite{pump2} and Y.Tserkovnyak et.al\cite{pumping},
and was confirmed experimentally by S.Mizukami et.al.\cite{mizukami}
In a spin Hall system, i.e. in a nonmagnetic semiconductor, 
Kato et.al\cite{kato} reported an observation
of a spin current by measuring optically the spin accumulation 
which appears as a result of spin currents at the edge of samples (\text{GaAs and InGaAs}). 
A critical issue in the observation of a spin current,
however, is that 
a spin current is not generally conserved and therefore
measuring spin accumulation 
does not necessarily indicate
the detection of a spin current, 
in sharp contrast to the case of charge.  
Non-conservation of spins is represented by a 
spin relaxation torque,  ${\mathbf{\cal{T}}}_{\text{s}}$,  which appears in the spin continuity equation.  
For a clear interpretation of experimental results on a spin current,
to understand the relaxation torque is essential.


The spin current means a flow of the spin angular momentum in general, 
and in metals conduction electrons carry  a spin current.
In insulators, 
there is no conduction electrons,
but there exists an other kind of carrier, namely, spin-waves,
which are  collective  motions  of  magnetic  moments.  
Experimentally, 
a spin-wave spin current,
a spin current carried by spin-waves
has already been established as a physical quantity.
Kajiwara\cite{spinwave} et.al have shown that 
a spin-wave spin current in an insulator can be generated and detected 
using direct and inverse spin-Hall effects.
They have revealed the  conversion of an electric signal into spin-waves, 
and its subsequent transmission through an insulator over macroscopic distance.
The spin-wave spin current has a novel feature\cite{spinwave}; 
this current  persists for much greater distance than the conduction electron spin current in metals, 
which disappears within very short distance (typically micrometers). 
For example in the magnetic insulator ({$\text{Y}_{3}\text{Fe}_{5}\text{O}_{12}$}), 
the spin-wave decay length can be several centimetres.   

In contrast to the experimental development,
theoretical studies so far of magnon transports 
are not enough to explain the experimental result on bulk systems. 
Meier and Loss\cite{meier} have investigated the magnon transport 
in both ferromagnetic and antiferromagnetic materials
and found that the spin conductance is quantized in the units of order $(g\mu_{\text{B}})^2/h$
in the antiferromagnetic isotropic spin-1/2 chains 
($g$ is the gyromagnetic ratio, $\mu_{\text{B}}$ is the Bohr magneton and $h$ is Planck constant).
Wang et.al\cite{china} have investigated a spin current carried by magnons
and derived a Landauer-B$\ddot{\text{u}}$ttiker-type formula for spin current transports.
They have also studied the magnon transport properties of a two-level magnon quantum dot 
in the presence of the magnon-magnon scattering 
and obtained the nonlinear spin current as a function of the magnetochemical potential.
These theoretical studies of magnon transports are
limited to mesoscopic systems.
From the viewpoint of spintronics,
the magnon transport in a macroscopic scale is an urgent and important subject.

In this paper, we focus on three dimensional ferromagnetic insulators.
The magnon source term, $\mathcal{T}_{\text{m}}$, 
arising from a time-dependent transverse magnetic field
is derived microscopically 
through Heisenberg's equation of motion.
We evaluate it by using Green's function
without relying on the phenomenological equation,
Landau-Lifshitz-Gilbert equation.
This is the main aim of this paper. 
The emergence of this term is 
in sharp contrast to a charge current
and represents the non-conservation of the magnon number.

This paper is structured as follows.
In \S  \ref{subsec:def},
we represent spin variables of a ferromagnetic insulator
by boson creation/annihilation operators via Holstein-Primakoff transformation.
We then apply a time-dependent transverse magnetic field. 
This magnetic field generates  the magnon source term,
which breaks the magnon conservation law
in the spin continuity equation.
In \S  \ref{subsec:source} and \S  \ref{subsec:frequency},
by evaluating the magnon source  term  at the low-temperature limit,
the dependence of the magnon source term on the angular frequency of a transverse magnetic field is calculated.
The magnon source term has a resonating behavior
when the angular frequency of an external transverse magnetic field is tuned.
In \S  \ref{subsec:temp}, the temperature dependence 
of the magnon source term is argued.
Through the analogy with the usual conduction electrons' spin pumping effect, 
the possibility for the magnon pumping is discussed in \S  \ref{sec:pumping}.

\section{Magnon Source Term }
\label{sec:mag}

\subsection{Definition}
\label{subsec:def}

We consider a ferromagnetic Heisenberg model in three dimensions.
It reduces to a free boson system via Holstein-Primakoff transformation
if we approximate the spin as ($S$ : the length of a spin) ; 
$S^{{z}}=S-a^{\dagger }a,  
(S^{+})^{\dagger }= S^{-}=(2S)^{1/2}a^{\dagger }[1-{a^{\dagger }a}/(2S)]^{1/2}\simeq (2S)^{1/2}a^{\dagger }$.
Here operators $a^{\dagger }/a $ are magnon creation/annihilation operators
satisfying the bosonic commutation relation.
Therefore in the continuous limit, a three-dimensional ferromagnetic insulator with an external 
magnetic field along the quantization ($z$) axis is described 
at low magnon density limit as
\begin{equation}
\begin{split}
{\cal{H}}_{\text{0}} &= {\cal{H}}_{\text{Heisen}} + {\cal{H}}_{\text{B}}  \\
&=  \int d^3x a^{\dagger }(\mathbf{x}t) \Big(-\frac{\hbar ^2 {\mathbf{\nabla }}^2 }{2m_{\text{mag}}} + g\mu _{\text{B}}\tilde B \Big)  a(\mathbf{x}t).
\;
\end{split}
\label{eqn:A0}
\end{equation}
Here $m_\text{mag} $ is the effective mass of a magnon
and it is represented by  a ferromagnetic exchange coupling constant in the discrete model, $ J $, and 
the (square) lattice constant, $ a_0 $, as
$  \hbar ^2/(2m_{\rm{mag}}) = 2JS{a_0}^2  $.
In eq.(\ref{eqn:A0}),
$\tilde B$ is a constant external magnetic field along the quantization axis ($z$-axis),
$g$ is $g$-factor and $\mu _{\text{B}}$ is  Bohr magneton.
From now on including $g$-factor and Bohr magneton, we write an external magnetic field  
as $g\mu _{\text{B}}\tilde B \equiv B.$
We then apply a time-dependent transverse magnetic field with an angular frequency, $ \Omega  $,
and a constant field strength,  $\Gamma _0$, 
to $x$-axis as
$\Gamma(t)=\Gamma _0 \rm{cos}\Omega t$.
\begin{equation}
\begin{split}
 V_{\Gamma }(t)&= \Gamma(t) \int d^3x S^x (\mathbf{x}) \\
            &\simeq  \Gamma(t) \int d^3x (\frac{S}{2})^{1/2} \bigg[ a(\mathbf{x}t)+a^{\dagger }(\mathbf{x}t)\bigg ]. \\
 \;
\end{split}
\label{eqn:A1}
\end{equation}
The total Hamiltonian is ${\cal{H}}={\cal{H}}_{\text{0}}+V_{\Gamma }(t)$.

The magnon density, $\rho_{\text{m}} (\mathbf{x})$, of the system is defined as the expectation value
of the number operator of magnons
\begin{equation}
\rho_{\text{m}} (\mathbf{x},t)\equiv \langle a^{\dagger }(\mathbf{x},t)a(\mathbf{x},t)\rangle.\;
\label{eqn:A3}
\end{equation}
Through Heisenberg's equation of motion,
the magnon current density, $\mathbf{ j_{\text{m}}} $, and the magnon source term, $ \mathcal{T}_{\text{m}}$, are defined as 
\begin{equation}
\begin{split}
\frac{\partial  \rho_{\text{m}} }{\partial  t} &= \frac{1}{i\hbar }[\rho_{\text{m}}, \cal{H}] \\
                                             &= -\mathbf{\nabla} \cdot \mathbf{ j_{\text{m}}} + \mathcal{T}_{\text{m}}.\;
\label{eqn:A4}
\end{split}
\end{equation}
Here the magnon current density 
arises from the free part;
$  [\rho_{\text{m}}, {\cal{H}}_{\text{0}}] / (i\hbar) $ 
= $ -\mathbf{\nabla} \cdot \mathbf{ j_{\text{m}}}$.
It reads
\begin{equation}
\begin{split}
j_{\text{m}}^{\mu } (\mathbf{x},t) &= \frac{\hbar }{m_{\text{mag}}}\text{Re} \ \Big[i<(\partial_\mu a^{\dagger }(\mathbf{x}t))a(\mathbf{x}t) >\Big] ,\;
\end{split}
\label{eqn:A5-0}
\end{equation}
where $ \mu $ is a direction for a magnon current to flow ($\mu = x,y,z  $).
The magnon source term, which represents the breaking of magnon conservation, 
arises from a transverse magnetic field as
$ [\rho_{\text{m}}, V_{\Gamma }]/ (i\hbar) $ $\equiv $ $\mathcal{T}_{\text{m}}$, i.e.,
\begin{equation}
\begin{split}
\mathcal{T}_{\text{m}}(t) = -\frac{(2S)^{1/2}}{\hbar }  \  \text{Im}\Big<\Gamma (t)  a(\mathbf{x}t) \Big>. \\
 \;        
\end{split}
\label{eqn:A5}
\end{equation}
From now on, we treat $V_{\Gamma }(t)$ as a perturbation
(i.e. a weak transverse magnetic field)
and study the effects of a time-dependent transverse magnetic field 
to the magnon source term. 

\subsection{Evaluation } 
\label{subsec:source}            
 
Through the standard procedure of the Keldysh (or contour-ordered) Green's function,\cite{kita,ramer,kamenev} 
the Langreth method,\cite{tatara,haug}
the magnon source term  is evaluated 
(see also APPENDIX \ref{sec:langreth}) as 
\begin{equation}
\begin{split}
<\Gamma (t) a(t)> =  \int d^3x' \int dt' \Gamma (t) \Gamma (t') (\frac{S}{2})^{1/2}G^{\text{r}}(t,t')  \\
 \ \ \ \ \ \ \ \ \ \ \ \ \ \ \ \ \   +O(\Gamma ^3).
 \;                               
\end{split}
\label{eqn:sourceT}
\end{equation}
Here $G^{\text{r}}$ is the retarded Green's function
and we have neglected  terms which are third-order in $\Gamma$,
which is justified at the low magnon density regime.

The retarded Green's function is
$G^{\text{r}}(\mathbf{r} \mathbf{r'},tt')= (\hbar/V) \sum_{\mathbf{k}}{\int (d\omega/2\pi )
e^{i\mathbf{k}\cdot (\mathbf{r}-\mathbf{r'})-i\omega (t-t')}G^{\text{r}}_{\mathbf{k},\omega }}$,  and  $
\ G^{\text{r}}(\mathbf{k},\omega )=[{\hbar \omega -\omega _{\mathbf{k}}+i\hbar/(2\tau)  }]^{-1}.$ 
Here $V$ is a volume of the system.  
The lifetime $\tau$ represents the damping of spins
($\tau $ is inversely
proportional to the Gilbert damping parameter,\cite{tatara} $\alpha $).
The energy $\omega _{\mathbf{k}}$ corresponds to the free part, ${\cal{H}}_{\text{0}}$,
and therefore
$\omega _{\mathbf{k}} = Dk^2 + B$, $D\equiv \hbar ^2/(2m_{\text{mag}})$.
Then the magnon source term is calculated as
\begin{equation}
\begin{split}
{\cal{T}}_{\text{m}} &= \frac{\Gamma ^2 _0}{4\hbar V }
S \bigg[ \frac{\frac{\hbar }{2\tau }+(\hbar \Omega +B)\rm{sin}2\Omega t+\frac{\hbar }{2\tau }\rm{cos}2\Omega t}{(\hbar \Omega +B)^2+(\frac{\hbar }{2\tau })^2}  \\
  & \ \ \ \ \ \ \ \ \ \ \ \ \ +\frac{\frac{\hbar }{2\tau }+(\hbar \Omega -B)\rm{sin}2\Omega t+\frac{\hbar }{2\tau }\rm{cos}2\Omega t}{(\hbar \Omega -B)^2+(\frac{\hbar }{2\tau })^2} \bigg ].
\;
\end{split}
\label{eqn:B4}
\end{equation}
The time average of $\cal{T}_{\text{m}}$ becomes
\begin{equation}
\begin{split}
{\cal{\bar T}}_{\text{m}} &= \frac{\Gamma ^2 _0}{4\hbar V }
S  \bigg[\frac{\frac{\hbar }{2\tau }}{(\hbar \Omega +B)^2+(\frac{\hbar }{2\tau })^2}  
   + \frac{\frac{\hbar }{2\tau }}{(\hbar \Omega -B)^2+(\frac{\hbar }{2\tau })^2} \bigg].
\end{split}
\label{eqn:zero}
\end{equation}
It is clear that $\cal{\bar T}_{\text{m}}$ is  positive
(for finite temperature, see eq.(\ref{eqn:B5-2}) in \S  \ref{subsec:temp}).


\subsection{Resonance} 
\label{subsec:frequency}                 

\begin{figure}[H]
\begin{center}
\includegraphics[width=7cm,clip]{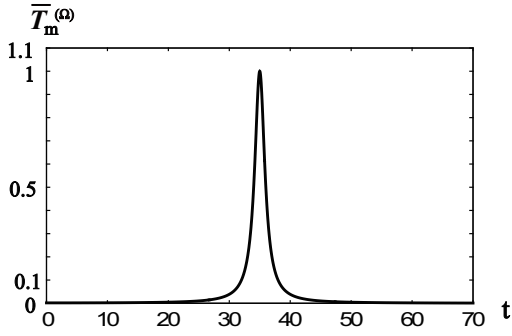}
\caption{
A graph of  ${\cal{\bar T}}_{\text{m}}^{(\Omega )} $, 
which represents the angular frequency dependence of the magnon source term. 
Parameters we have used are,
 $2\tau \Omega  \equiv t$, 
$\tau =2\times 10^{-6}[\text{s}]$,
$\tilde B=1[\text{G}]$,
$g=1$.
Therefore $2\tau B/\hbar $ is $ 35$.
This ${\cal{\bar T}}_{\text{m}}^{(\Omega )} $ has a sharp peak 
around $ t =35$.
This fact means that the magnon source term has a resonance structure
with  the applied transverse magnetic field.
\label{fig:T}}
\end{center}
\end{figure}

We define a dimensionless quantity
${\cal{\bar T}}_{\text{m}}^{(\Omega )}$ as
\begin{equation}
\begin{split}
{\cal{\bar T}}_{\text{m}}^{(\Omega )}&\equiv \frac{1}{(2\tau \Omega +\frac{2\tau B}{\hbar })^2+1}+
\frac{1}{(2\tau \Omega -\frac{2\tau B}{\hbar })^2+1}, \ \text{i.e.,}  \\
\cal{\bar T}_{\text{m}}&=\frac{\Gamma ^2 _0}{4\hbar V}  S \cdot   \frac{2\tau }{\hbar }{\cal{\bar T}}_{\text{m}}^{(\Omega )}. 
\end{split}
\label{eqn:omega}
\end{equation}
This shows that 
the magnon source term 
has a resonance structure
with a time-dependent transverse magnetic field
when the angular frequency is tuned as 
$ \Omega  =    B/ \hbar   $
(see Fig.\ref{fig:T}).
This resonance is useful for the enhancement of the magnon pumping.

\subsection{Temperature dependence} 
\label{subsec:temp}           

Let us look into the temperature dependence of the magnon source term.
To do this, we include the interaction of third-order in magnon operators.
Then $V_{\Gamma }(t) $ is rewritten as
\begin{equation}
\begin{split}
 V_{\Gamma }(t)&= \Gamma(t) \int d^3x S^x (\mathbf{x}) \\
            &\simeq   \Gamma(t) \int d^3x (\frac{S}{2})^{1/2} \bigg\{ a(\mathbf{x}t)+a^{\dagger }(\mathbf{x}t)  \\
&-\frac{1}{4S}\Big[ a^{\dagger }(\mathbf{x}t) a^{\dagger }(\mathbf{x}t)a(\mathbf{x}t)    + a^{\dagger }(\mathbf{x}t) a(\mathbf{x}t) a(\mathbf{x}t)  \Big]\bigg\}, \;
\end{split}
\label{eqn:koreda}
\end{equation}
and  the magnon source term  reads
\begin{equation}
\begin{split}
\mathcal{T}_{\text{m}}(t) &= -\frac{(2S)^{1/2}}{\hbar }  \  \text{Im}\bigg<\Gamma (t) \Big[ a(\mathbf{x}t) + \frac{a^{\dagger }(\mathbf{x}t)a^{\dagger }(\mathbf{x}t)a(\mathbf{x}t)}{4S}   \Big]\bigg>. \;        
\end{split}
\label{eqn:A5}
\end{equation}
Eq.(\ref{eqn:A5}) is calculated as
\begin{equation}
\begin{split}
<\Gamma (t) a(t)> &=  \int d^3x' \int dt' \Gamma (t) \Gamma (t')\Big[(\frac{S}{2})^{1/2}G^{\text{r}}(t,t')   \\
            &\ \ \ -\frac{i}{2(2S)^{1/2}} G^{\text{r}}(t,t') G^{\text{<}}(t',t')\Big]+O(\Gamma ^3),       
\;                               
\end{split}
\label{eqn:B1}
\end{equation}
\begin{equation}
\begin{split}
<\frac{\Gamma (t)}{4S} a^{\dagger }(t)a^{\dagger }(t)a(t)> &=   \frac{i}{2(2S)^{1/2}} \int d^3x' \int dt'       \\
  &\cdot    \Gamma (t) \Gamma (t')    G^{\text{a}}(t',t) G^{\text{<}}(t,t) +O(\Gamma ^3).      \;                                 
\end{split}
\label{eqn:B2}
\end{equation}
Here $G^{\text{a}}$ and $G^{\text{<}}$ are the advanced and lesser Green's functions, respectively.
Because we focus on the behavior of the magnon source term at low temperature that
we have neglected higher terms than the fourth-order in respect to magnon creation/annihilation operators.
The Fourier transform of the lesser Green's function satisfies,
$\ G^{\text{<}}(\mathbf{k},\omega )=-f_{\text{B}}(\omega _{\mathbf{k}})[G^{\text{a}}(\mathbf{k},\omega )-G^{\text{r}}(\mathbf{k},\omega )]$,
where  $G^{\text{a}}(\mathbf{k},\omega )=[{\hbar \omega -\omega _{\mathbf{k}}-i\hbar/(2\tau)  }]^{-1}   $.
Then $\cal{T}_{\text{m}}$ is calculated as 
\begin{equation}
\begin{split}
\cal{T}_{\text{m}} &= \frac{\Gamma ^2 _0}{4\hbar V}
\bigg[ S -\Big(1+\frac{1}{2^{3/2}}e^{-\beta B}\Big)e^{-\beta B}\Big(\frac{ k_{\rm B}T}{4\pi D}\Big)^{3/2} \bigg] \\ 
&\ \ \ \ \ \ \ \ \cdot  \bigg[ \frac{\frac{\hbar }{2\tau }+(\hbar \Omega +B)\rm{sin}2\Omega t+\frac{\hbar }{2\tau }\rm{cos}2\Omega t}{(\hbar \Omega +B)^2+(\frac{\hbar }{2\tau })^2}  \\
 &\ \ \ \ \ \ \ \ \ \ +\frac{\frac{\hbar }{2\tau }+(\hbar \Omega -B)\rm{sin}2\Omega t+\frac{\hbar }{2\tau }\rm{cos}2\Omega t}{(\hbar \Omega -B)^2+(\frac{\hbar }{2\tau })^2} \bigg ],
\;
\end{split}
\label{eqn:B4}
\end{equation}
where ${\beta}^{-1}$ is  $k_{\text{B}}T $ ($k_{\text{B}} :$ Boltzmann constant).
Here we have approximated the summation over the Bose distribution function as,
$ [(2\pi )^3/V]\sum_{\mathbf{k}}f_{\text{B}}(\omega _{\mathbf{k}})  = 2\pi \int_{-\infty }^{\infty } dk k^2[e^{\beta (Dk^2+B)} -1 ]^{-1} 
                                                                                               \simeq 2\pi \int_{-\infty }^{\infty } dk \ k^2 e^{-\beta (Dk^2+B)} [1+ e^{-\beta (Dk^2+B)}]$.                                      

The time average of $\cal{T}_{\text{m}}$ becomes
\begin{equation}
\begin{split}
\cal{\bar T}_{\text{m}} &= \frac{\Gamma ^2 _0}{4\hbar V}
\bigg[ S -\Big(1+\frac{1}{2^{3/2}}e^{-\beta B}\Big)e^{-\beta B}\Big(\frac{  k_{\rm B}T}{ 4\pi D}\Big)^{3/2}\bigg] \\
 & \ \ \   \cdot  \bigg[\frac{\frac{\hbar }{2\tau }}{(\hbar \Omega +B)^2+(\frac{\hbar }{2\tau })^2} 
+\frac{\frac{\hbar }{2\tau }}{(\hbar \Omega -B)^2+(\frac{\hbar }{2\tau })^2} \bigg] \\
&= \frac{\Gamma ^2 _0}{4\hbar V } \frac{2\tau }{\hbar }{\cal{\bar T}}_{\text{m}}^{(\Omega )}{\cal{\bar T}}_{\text{m}}^{(T )}, \\
\end{split}
\label{eqn:B5-2}
\end{equation}
where
\begin{equation}
\begin{split}
{\cal{\bar T}}_{\text{m}}^{(T )}&\equiv S -\Big(1+\frac{1}{2^{3/2}}e^{-\beta B}\Big)e^{-\beta B}\Big(\frac{ k_{\rm B}T}{4\pi D}\Big)^{3/2},
\end{split}
\label{eqn:B5-2-2}
\end{equation}
is the temperature-dependent part and 
 ${\cal{\bar T}}_{\text{m}}^{(\Omega )} $ is defined in eq.(\ref{eqn:omega}).
When the temperature gets higher,
the magnon source term decreases.
This means that quantum and thermal fluctuations 
act in the opposite way,
namely to increase and decrease the magnon source term, respectively.
It is clear that eq.(\ref{eqn:B5-2}) reduces at the zero temperature to  eq.(\ref{eqn:zero}).

\section{Magnon Pumping} 
\label{sec:pumping}                

The spin pumping is an effect widely used to create a spin current by use of the magnetization precession. \cite{takeuchi,mizukami, pump2,pump3,referee}
Experiments have been carried out in junctions of metallic ferromagnets and nonmagnetic metals.
According to the theory by
Y.Tserkovnyak and A.Brataas,\cite{pumping}
the spin current pumped through the junction reads
${\mathbf{I}}_{\text{s}}^{\text{pump}}= [\hbar /(4\pi)] [ A_{\text{r}}\mathbf{m}\times (d{\mathbf{m}}/dt) -A_{\text{i}}  (d{\mathbf{m}}/dt) ]$,
where $\mathbf{m}$ is the magnetization direction of a localized spin and  
$A_{\text{r}},  A_{\text{i}}$ are the interface parameters.
This result is understood by considering the spin continuity equation, 
$\nabla\cdot \mathbf{j}_{\text{s}}=-(\partial {\mathbf{m}} / \partial t) +{\mathbf{\cal{T}}}_{\text{s}}$, where $\mathbf{j}_{\text{s}}$ is the spin current density and 
${\mathbf{\cal{T}}}_{\text{s}}$ is the spin relaxation torque.
In fact, the spin continuity equation indicates that a spin current is generated when the magnetization is dynamic and/or when ${\mathbf{\cal{T}}}_{\text{s}}$ is finite.
It has been well-known \cite{tatara} that
the main term of spin relaxation torque, ${\mathbf{\cal{T}}}_{\text{s}}$, has the form 
${\mathbf{\cal{T}}}_{\text{s}}\propto \mathbf{m}\times (d\mathbf{m}/dt) $
 in metals.
 As Brataas et al.\cite{battery} have pointed out,
 this torque  ${\mathbf{\cal{T}}}_{\text{s}} $  is equivalent to the maximum spin current that can be drawn from the spin battery. 
The pumping formula for ${\mathbf{I}}_{\text{s}}^{\text{pump}}$ in metals  is thus understood from the spin continuity equation.
The relaxation torque plays an essential role in spin pumping in metals, and it is expected to be  dominant also in the spin pumping in insulating ferromagnets.
Our calculation of the relaxation torque term (i.e. the magnon source term)  thus describes the spin pumping effect in insulators.

We have revealed 
that the magnon source term has a sharp peak around $ \Omega  =  B/\hbar $,
as a result of the resonance with a time-dependent transverse magnetic field 
when the angular frequency is appropriately adjusted as 
$ \Omega  =  B/\hbar  $.
This fact is useful to enhance the magnon pumping effect
because the external magnetic field, $B$, and 
the angular frequency of a transverse magnetic field, $\Omega $, is under our control.


Experimentally, the magnon pumping effect we have discussed 
can be identified by observing the temperature dependence  of 
the pumped magnon current when $B$ is zero;
$ ( \text{the pumped magnon current} ) \propto   {\cal{\bar T}}_{\text{m}}^{(T )}  \propto   (\text{A}_1 - \text{A}_2 T^{3/2} ) $,
where $\text{A}_1$ and $\text{A}_2$ are constants.

\begin{figure}[h]                          
\begin{center}                              
\includegraphics[width=5cm,clip]{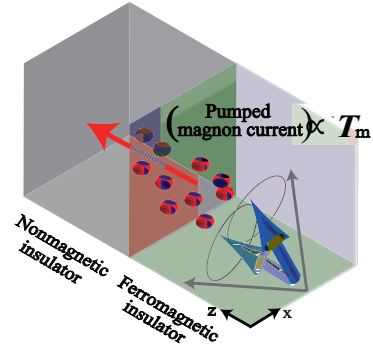}  
\caption{(Color online)                     
A schematic picture of the magnon pumping
by a time-dependent transverse magnetic field. 
Thick and short solid arrows represent localized spins,
and thin and long ones external magnetic fields.  
Circles represent magnons.
By way of resonance with a time-dependent transverse magnetic field, 
the magnon source term is enhanced,
and an enhanced magnon current is pumped    
from the ferromagnetic insulator  to the adjacent nonmagnetic insulator.
\label{fig:pump}}                           
\end{center}                                
\end{figure}                                

\section{Summary }
\label{sum}
We have studied theoretically the magnon source term which represents 
the breaking of the magnon conservation law.
We have revealed 
that the magnon source term has a resonance structure 
with an external time-dependent transverse magnetic field
when the angular frequency of the applied magnetic field is tuned.
This fact will be useful to enhance the magnon pumping effect in insulators.
This magnon pumping effect is a new method for a generation of 
a magnon current (spin-wave spin current) without the gradient of 
an external magnetic field.

\begin{acknowledgments}
The author (K.N) would like to thank 
K.Totsuka, M.Oshikawa, Y.Korai and K.Taguchi
for useful comments and discussion.
One of the author(G.T) is supported by a Grant-in-Aid
for Scientific Research in Priority Areas, $"$Creation and control
of spin current$"$ under Grant No. 1948027, 
and a Grant-in-Aid for Scientific Research (B) (Grant No. 22340104).
\end{acknowledgments}

\appendix 
\section{Calculation of Equation.(\ref{eqn:sourceT})}
\label{sec:langreth}

\begin{figure}[h]
\begin{center}
\includegraphics[width=5cm,clip]{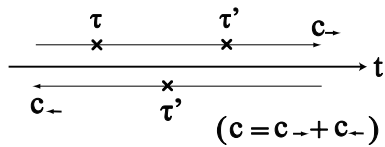}
\caption{Keldysh contour, $c$. We have taken $\tau $ on forward path, ${\text{c}}_{\rightarrow }$.
Even when $\tau $ is located  on backward path, ${\text{c}}_{\leftarrow }$, 
the result of this calculation is invariant
because each Green's function, $G^{\text{r}} $, $G^{\text{a}} $, $G^{\text{<}} $, $G^{\text{>}} $ (the greater Green's function ),
is not independent; they obey, 
$G^{\text{r}}  - $ $G^{\text{a}} $ $ = G^{\text{>}} $  $ - G^{\text{<}} $. 
Both the forward and backward paths are actually on the real axis 
but shifted slightly upwards and downwards, respectively, to distinguish them clearly.
\label{fig:langreth}}
\end{center}
\end{figure}

In this section, we briefly show the Langreth method,
which is useful to evaluate the perturbation expansion of 
the Keldysh ( or contour-ordered) Green's function.
For simplicity here, we evaluate the magnon source term at zero temperature, eq.(\ref{eqn:sourceT}),  as an example;

\begin{equation}
\begin{split}
V_{\Gamma }(t) &\simeq  \Gamma(t) \int d^3x (\frac{S}{2})^{1/2}  \bigg[ a(\mathbf{x}t)+a^{\dagger }(\mathbf{x}t) \bigg], \\
\mathcal{T}_{\text{m}}(t) &= -\frac{(2S)^{1/2}}{\hbar }  \  \text{Im} \bigg<\Gamma (t) a(\mathbf{x}t) \bigg> .\\
 \;              
\end{split}
\label{eqn:appe2}
\end{equation}
We have only to estimate $<a(\mathbf{x}t)>$.
It is evaluated as
\begin{equation}
\begin{split}
<a(\tau )> &= \bigg<\text{T}_{\text{c}} \ a(\tau ) \text{exp}\Big[-i\int_{\text{c}} d\tau ' V_{\Gamma }(\tau ')  \Big]  \bigg>   \\         
           & \simeq   -i(\frac{S}{2})^{1/2}\int d^3x'  \\
           & \ \ \  \cdot  \int_{\text{c}} d\tau ' \Gamma (\tau ')\Big<\text{T}_{\text{c}} \ a(\mathbf{x}\tau)a^{\dagger }(\mathbf{x'}\tau ')\Big> \\
            & \equiv     -i(\frac{S}{2})^{1/2}\int d^3x'   {\cal{I}}    .\;
\end{split}
\label{eqn:appe3}
\end{equation}
Here $\text{T}_{\text{c}}$ is the path-ordering operator defined 
on the Keldysh contour, $\text{c}$ (see Fig.\ref{fig:langreth}).
We express  the Keldysh contour as a sum of the forward path,
$c_{\rightarrow }$, and the backward path, $c_{\leftarrow }$;
$ c=c_{\rightarrow} + c_{\leftarrow} $.
The integral on the  Keldysh contour of eq.(\ref{eqn:appe3}), $  {\cal{I}} $,  is executed 
by taking an identity into account 
\begin{equation}
\begin{split}
\int_{{\text{c}}} d\tau^{{\text{c}}} &= \int_{{\text{c}}_{\rightarrow }} d\tau^{{{\rightarrow }}}
+ \int_{{\text{c}}_{\leftarrow  }} d\tau^{{{\leftarrow }}}, \\
  \;
\end{split}
\label{eqn:kel}
\end{equation}
as
\begin{equation}
\begin{split}
{\cal{I}} = i\int_{-\infty }^{\infty } d\tau' \Gamma (\tau ') \Big[ G^{\text{t}}(\tau ,\tau ') -  G^{\text{<}}(\tau ,\tau ') \Big  ] . \;
\end{split}
\label{eqn:appe5}
\end{equation}
Here $G^{\text{t}} $ is the time-ordered Green's function. 
By using the relation,
$ G^{\text{r}}(t,t') = G^{\text{t}}(t,t') - G^{\text{<}}(t,t') $,
we obtain eq.(\ref{eqn:sourceT}).

\end{document}